\documentclass[12pt,a4paper]{article}

\usepackage[utf8]{inputenc}
\usepackage[T1]{fontenc}
\usepackage[english]{babel}
\usepackage{amsmath,amssymb}
\usepackage{graphicx}
\usepackage{booktabs}
\usepackage{array}
\usepackage{hyperref}
\usepackage{geometry}
\usepackage{caption}
\usepackage{natbib}
\usepackage{microtype}
\usepackage{float}
\usepackage{xcolor}
\usepackage{doi}
\hypersetup{
  colorlinks=true,
  linkcolor=blue!60!black,
  citecolor=blue!60!black,
  urlcolor=blue!60!black
}

\title{Comet 1P/Halley Completes 15 Orbits in 1,151 Years:\\
Commensurability with the Solar System Quasi-Period\\
and \\Evidence for Jupiter-Saturn Dynamical Coupling}

\author{Carlos \textsc{Baiget Orts}\thanks{Correspondence:
  \href{mailto:asinfreedom@gmail.com}{asinfreedom@gmail.com}.
  ORCID: \href{https://orcid.org/0009-0000-6725-5188}{0009-0000-6725-5188}.
  Family name: Baiget Orts.}\\
  \small\textit{Independent researcher, Valencia, Spain}}

\date{}

\begin{document}

\maketitle

\begin{abstract}
\noindent
I investigate whether comet 1P/Halley participates in the
1{,}151-year planetary quasi-period $T^{*}$ identified in
\citet{BaigetOrts2026a}.
Using historical perihelion records spanning 2{,}225~years
(30~apparitions, 239~BCE to 1986~CE), I find that Halley's mean orbital period $\bar{P} = 76.713$~yr satisfies $T^{*}/\bar{P} = 15.004$, yielding an angular residue of $+1.43^\circ$
--- the smallest of any Solar System body examined, including all seven
planets that participate in $T^{*}$ (Mercury, Venus, Earth, Mars,
Jupiter, Saturn, and Neptune; $p = 0.009$).
No other Halley-type comet participates: all examined HTCs exhibit
residues of $80^\circ$--$130^\circ$, comparable to Uranus
($108^\circ$), the sole planetary non-participant.
Four independent statistical tests establish that Jupiter and Saturn
couple to Halley's orbital period through distinct mechanisms.
Jupiter acts through \textit{phase-dependent modulation}: its angular
position at each perihelion predicts the period deviation
($p = 0.027$--$0.04$, three methods).
Saturn acts through \textit{distance-amplitude modulation}: closer
approaches produce larger deviations regardless of sign
($r = -0.496$, $p = 0.007$), specific to Saturn's actual orbital
phase (random-phase control $p = 0.133$).
After 15~orbits, the cumulative period deviation is only $9.4\%$ of
the random-walk expectation --- direct evidence of coherent
perturbation cancellation over one $T^{*}$ cycle.
The orbit-to-orbit chaos ($\tau_L \approx 70$~yr) and the long-term
mean stability are not contradictory: the same Jupiter--Saturn forces
that cause individual-orbit variability cancel coherently over the
$T^{*}$ baseline, anchoring the mean period at the millennium scale.


\end{abstract}
\newpage

\section{Introduction}\label{sec:intro}

In \citet{BaigetOrts2026a}, I reported the detection of a multi-planet quasi-commensurability in the Solar System: the interval $T^{*} = 420{,}403$~days ($\approx 1{,}151$~years)
minimizes a global similarity metric applied to the heliocentric
ecliptic longitudes of seven planets --- Mercury, Venus, Earth,
Mars, Jupiter, Saturn, and Neptune. At this interval, the mean
angular displacement of all seven planets from their positions
$T^{*}$~days earlier is only $13.4^\circ$, with a standard deviation
of $0.65^\circ$ sustained over a century-long comparison window.
Neptune participates with remarkable precision (angular residue
$-5.2^\circ$), while Uranus---the only planet known to have suffered a catastrophic giant impact that tilted its axis to $98^\circ$ \citep{Kegerreis2018}---does not (residue $-108.3^\circ$). The quasi-commensurability thus encompasses all planets in the Solar System except Uranus.

This result raises a natural question: do any other bodies in the Solar System participate in this quasi-commensurability? The present work examines this question for periodic comets, focusing on Halley-type comets (HTCs), which have orbital periods between 20 and 200~years and whose dynamics are dominated by perturbations from Jupiter and Saturn---the same planets that define the structure of $T^{*}$.

The prototype of this class, comet 1P/Halley, has been observed at
30~perihelion passages spanning over two millennia, providing a
uniquely long baseline for computing mean orbital parameters.
Its individual orbital period varies between 74.42 and 79.25~yr
\citep{Yeomans1986}, driven by gravitational perturbations from the
giant planets, non-gravitational forces from cometary outgassing,
and occasional close planetary approaches
\citep{ChirikovVecheslavov1989, MunozGutierrez2015}.
The orbit is formally chaotic, with a Lyapunov time of
approximately 70~yr \citep{MunozGutierrez2015} --- shorter than one
orbital period.
This raises a question that motivates the present work: if the orbit
is chaotic on the timescale of a single revolution, can its long-term
mean period carry a coherent dynamical signature of the planetary
quasi-period $T^{*}$?
The answer, as the following analysis shows, is yes.

\section{Data and Methods}\label{sec:data}

\subsection{Perihelion dates of 1P/Halley}\label{subsec:halley_data}

I adopt the historical perihelion dates compiled by \citet{Yeomans1986}, supplemented by the observed 1986 perihelion (February~9, 1986). The complete dataset comprises 30~apparitions from 239~BCE to 1986~CE (Table~\ref{tab:halley_perihelia}). Dates prior to 1582~CE are in the Julian calendar; dates from 1607~CE onward are in the Gregorian calendar.

The mean orbital period is computed as:
\begin{equation}\label{eq:mean_period}
\bar{P} = \frac{t_{N} - t_{1}}{N - 1}
\end{equation}
where $t_{1}$ and $t_{N}$ are the first and last perihelion dates, and $N = 30$ is the number of apparitions, yielding $N - 1 = 29$ complete orbital periods.

\subsection{Commensurability metric}\label{subsec:metric}

For a body with mean period $\bar{P}$, the commensurability with $T^{*}$ is quantified by the angular residue:
\begin{equation}\label{eq:residue}
\Delta\theta = \left(\frac{T^{*}}{\bar{P}} - \mathrm{nint}\!\left(\frac{T^{*}}{\bar{P}}\right)\right) \times 360^\circ
\end{equation}
where $\mathrm{nint}(x)$ denotes the nearest integer to $x$. A body with $|\Delta\theta| \approx 0^\circ$ completes a near-integer number of orbits in $T^{*}$~years; a body with $|\Delta\theta| \gg 0^\circ$ does not.

\subsection{Statistical significance}\label{subsec:statistics}

To assess whether the observed commensurability could arise by chance, I perform a Monte Carlo simulation, generating $10^{7}$ random periods drawn uniformly from the interval $[74, 80]$~yr (encompassing Halley's observed range) and computing the angular residue for each. The $p$-value is the fraction of random periods producing a residue $|\Delta\theta|$ equal to or smaller than the observed value.

\subsection{Halley-type comet survey}\label{subsec:htc_survey}
To determine whether the commensurability is generic among HTCs, I analyze
the three HTCs with observational baselines exceeding one century and at
least three well-determined apparitions: 12P/Pons--Brooks (6~confirmed apparitions, 1385--2024,
of which 4 are well-determined; \citealt{Meyer2021}), 55P/Tempel--Tuttle (5~confirmed apparitions, 1366--1998;
33~computed returns, 901--1998), and 27P/Crommelin (4~well-determined
apparitions, 1928--2011). HTCs with shorter baselines or fewer secure
apparitions are excluded, as the resulting mean periods would carry
uncertainties comparable to the commensurability window being tested.
For each, $\bar{P}$ and $\Delta\theta$ are computed following the same
procedure.

\section{Results}\label{sec:results}

\subsection{1P/Halley: the primary result}\label{subsec:halley_result}

The total time span from the first recorded perihelion (239~BCE, approximately May~15) to the last (1986~CE, February~9) is 2,225~years. Dividing by 29~orbital periods yields:
\begin{equation}
\bar{P} = \frac{2224.68}{29} = 76.713 \text{~yr}
\end{equation}

The ratio with the 1151-year quasi-period is:
\begin{equation}
\frac{T^{*}}{\bar{P}} = \frac{1151}{76.713} = 15.004
\end{equation}

The angular residue is:
\begin{equation}
\Delta\theta = 0.004 \times 360^\circ = +1.43^\circ
\end{equation}

This corresponds to a deviation of only $0.020$~yr ($\approx 7.4$~days) from the exactly commensurable period $T^{*}/15 = 76.7333$~yr.

\begin{figure}[htbp]
  \centering
  \includegraphics[width=0.85\textwidth]{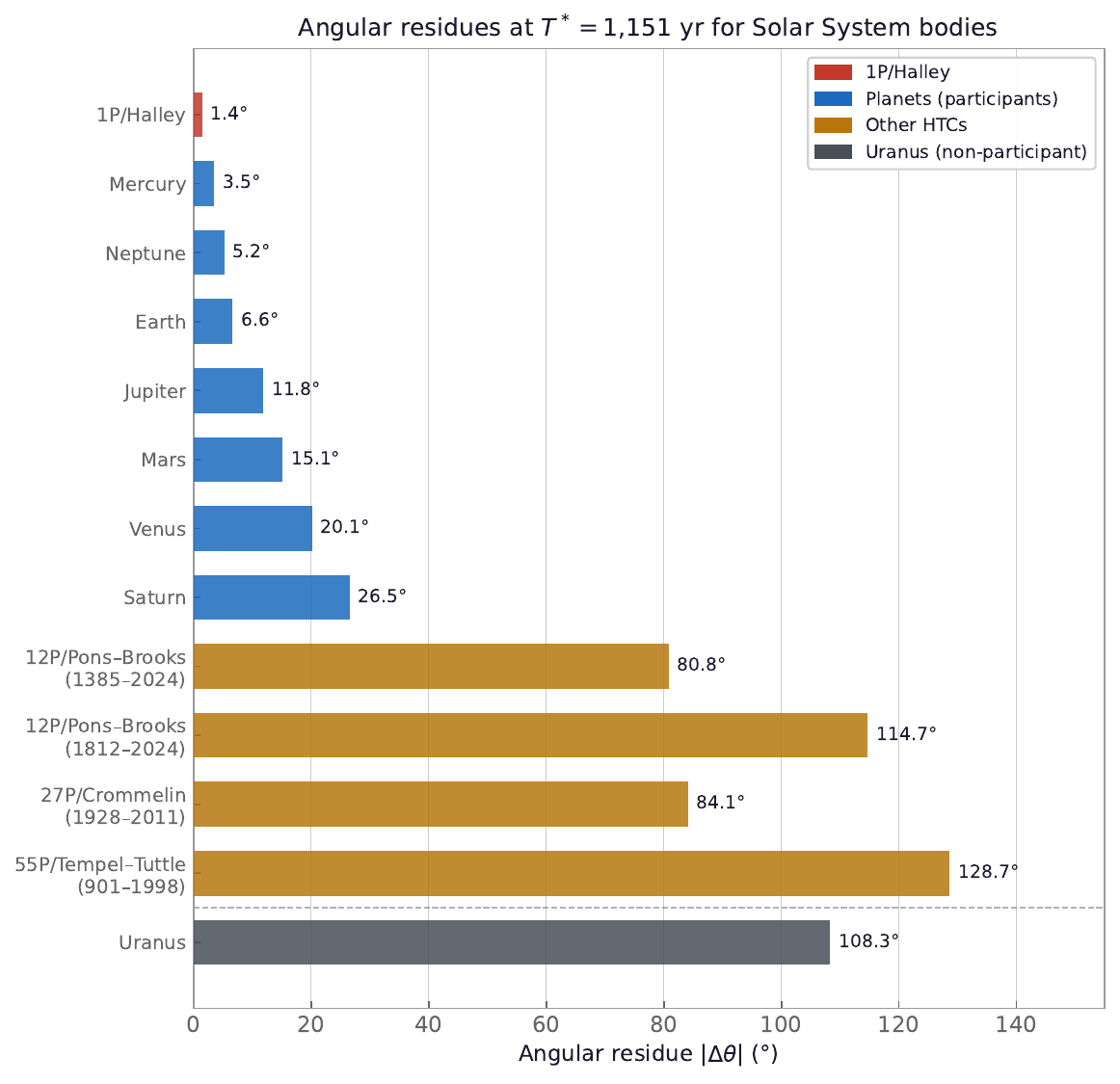}
  \caption{Angular residues $|\Delta\theta|$ at $T^{*} = 1{,}151$~yr
    for Solar System bodies.
    Planets are shown in blue, comet 1P/Halley in red, and
    other Halley-type comets in orange.
    Uranus (grey, dashed border) is shown separately as the sole
    non-participant among the planets.
    Halley's residue ($+1.43^\circ$) is the smallest of any body
    examined, smaller even than Mercury ($3.5^\circ$) and Neptune
    ($5.2^\circ$), while all other HTCs cluster between
    $80^\circ$ and $130^\circ$.}
  \label{fig:residues_comparison}
\end{figure}

Table~\ref{tab:residues} compares this residue with those of the planets from \citet{BaigetOrts2026a}. Halley's residue is the smallest of any body analyzed, including
all seven participating planets.
Figure~\ref{fig:residues_comparison} displays these residues graphically alongside those of the other Halley-type comets surveyed in Section~\ref{subsec:other_htcs}.

\begin{table}[htbp]
\centering
\caption{Angular residues at $T^{*} = 1{,}151$~yr for Solar System bodies.}
\label{tab:residues}
\begin{tabular}{lrrrr}
\toprule
Body & $\bar{P}$ (yr) & $T^{*}/\bar{P}$ & $N$ & $\Delta\theta$ ($^\circ$) \\
\midrule
\textbf{1P/Halley}  & 76.713  & 15.004   & 15   & $\mathbf{+1.4}$ \\
Mercury              & 0.2409  & 4778.99  & 4779 & $-3.5$ \\
Neptune              & 164.77  & 6.986    & 7    & $-5.2$ \\
Earth                & 1.0000  & 1150.98  & 1151 & $-6.6$ \\
Jupiter              & 11.862  & 97.03    & 97   & $+11.8$ \\
Mars                 & 1.8808  & 611.96   & 612  & $-15.1$ \\
Venus                & 0.6152  & 1870.94  & 1871 & $-20.1$ \\
Saturn               & 29.457  & 39.07    & 39   & $+26.5$ \\
\midrule
Uranus               & 84.02   & 13.699   & 14   & $-108.3$ \\
\bottomrule
\end{tabular}

\smallskip
\small\textit{Note.} Bodies are sorted by $|\Delta\theta|$.
Planetary residues are the arithmetic sidereal residues
$\Delta\theta = (\mathrm{frac}(T^{*}/P)) \times 360^\circ$
using DE441 sidereal periods from \citet{BaigetOrts2026a};
they differ slightly from the empirical metric scores reported
there, which combine mean and standard deviation of the angular
displacement series.
Halley's residue is computed from the mean of 29~observed orbital
periods using precise Julian Day perihelion dates.
Uranus is listed separately as it does not participate in the
quasi-commensurability.

\end{table}

\subsection{Variability and convergence}\label{subsec:variability}

The 29~individual orbital periods range from 74.4 to 79.3~yr, with a standard deviation of 1.25~yr (Table~\ref{tab:halley_perihelia}). This variability is driven primarily by gravitational perturbations from Jupiter and Saturn at each perihelion passage \citep{Yeomans1986}, with secondary contributions from non-gravitational forces due to cometary outgassing \citep{ChirikovVecheslavov1989}.

\begin{figure}[htbp]
  \centering
  \includegraphics[width=0.85\textwidth]{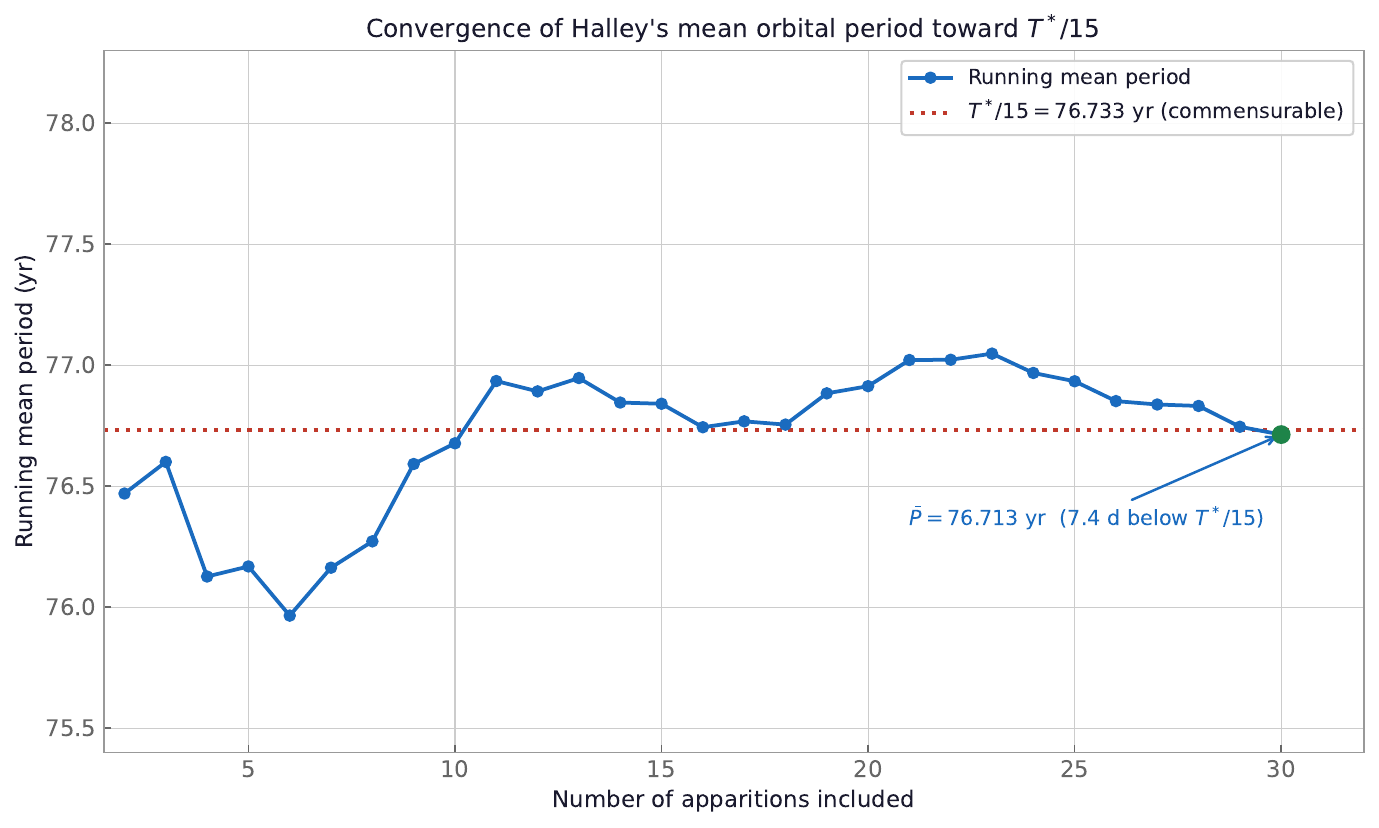}
  \caption{Convergence of the running mean orbital period of comet
    1P/Halley toward the exactly commensurable value $T^{*}/15 =
    76.733$~yr (red dotted line), as successive perihelion passages
    are included.
    The final observed mean $\bar{P} = 76.713$~yr lies only 7.4~days
    below $T^{*}/15$ after 29~observed orbital periods spanning
    2{,}225~years.}
  \label{fig:periods_convergence}
\end{figure}

Despite this substantial orbit-to-orbit variability ($\pm 2$~yr), the perturbations cancel over the 29-orbit baseline with a precision of $1.43^\circ$---implying that the perturbations are not random but correlated with the planetary configuration, which itself recurs every $T^{*}$~years. Figure~\ref{fig:periods_convergence} shows the individual periods and the convergence of the running mean toward $T^{*}/15$.

\begin{table}[htbp]
\centering
\caption{Perihelion dates and orbital periods of 1P/Halley.}
\label{tab:halley_perihelia}
\begin{tabular}{rrr}
\toprule
App. & Year & $P_{i}$ (yr) \\
\midrule
1  & $-$239 & ---  \\
2  & $-$163 & 76   \\
3  & $-$86  & 77   \\
4  & $-$11  & 75   \\
5  & 66     & 77   \\
6  & 141    & 75   \\
7  & 218    & 77   \\
8  & 295    & 77   \\
9  & 374    & 79   \\
10 & 451    & 77   \\
11 & 530    & 79   \\
12 & 607    & 77   \\
13 & 684    & 77   \\
14 & 760    & 76   \\
15 & 837    & 77   \\
16 & 912    & 75   \\
17 & 989    & 77   \\
18 & 1066   & 77   \\
19 & 1145   & 79   \\
20 & 1222   & 77   \\
21 & 1301   & 79   \\
22 & 1378   & 77   \\
23 & 1456   & 78   \\
24 & 1531   & 75   \\
25 & 1607   & 76   \\
26 & 1682   & 75   \\
27 & 1759   & 77   \\
28 & 1835   & 76   \\
29 & 1910   & 75   \\
30 & 1986   & 76   \\
\bottomrule
\end{tabular}

\smallskip
\small\textit{Note.} Perihelion years are approximate to the nearest year.
Years prior to 1582~CE follow the Julian calendar; from 1607~CE onward
the Gregorian calendar is used.
Periods prior to 837~CE carry uncertainties of order months due to
limited historical records and non-gravitational modeling \citep{Yeomans1986}.
Individual periods are rounded to the nearest year; the range and
standard deviation are computed from precise Julian Day dates.
The mean period is $\bar{P} = 76.713$~yr; the standard deviation of
the 29~individual periods is $\sigma = 1.25$~yr, with a range of
$[74.4, 79.3]$~yr.

\end{table}

\subsection{Statistical significance}\label{subsec:pvalue}

The Monte Carlo test (Section~\ref{subsec:statistics}) yields a $p$-value of $0.036$ when drawing random periods uniformly from $[74, 80]$~yr and computing the mean of 29 samples: only $3.6\%$ of such synthetic comets produce a residue $|\Delta\theta| \leq 1.43^\circ$. For a single random period in the same range, the $p$-value is $0.007$.

However, the most stringent test accounts for the joint coincidence that $T^{*}$ is simultaneously optimal for the planets \textit{and} a near-exact multiple of Halley's mean period. A period scan over all candidate cycles $T = 100$--$2{,}000$~yr shows that $T^{*} = 1{,}151$ ranks first in the planetary metric, 16th for Halley alone, and \textit{first in the joint planetary-plus-Halley metric}, with a score more than double that of the second-best candidate. The probability that the best planetary cycle also falls in the top 16 for Halley by chance is $16/1901 = 0.84\%$. A Monte Carlo test generating $10^{5}$ random comet periods in $[20, 200]$~yr and searching for the best joint score across all candidate cycles confirms this: only $0.86\%$ of random comets achieve a joint score as good as the observed one ($p = 0.009$).

\subsection{Absence of commensurability in other HTCs}\label{subsec:other_htcs}

Table~\ref{tab:htc_survey} summarizes the results of the HTC survey. No other Halley-type comet with multiple observed apparitions exhibits commensurability with $T^{*}$.

\begin{table}[htbp]
\centering
\caption{Commensurability test for Halley-type comets with multiple apparitions.}
\label{tab:htc_survey}
\begin{tabular}{lrrcrr}
\toprule
Comet & $\bar{P}$ (yr) & $N_\mathrm{orb}$ & Span & $N$ & $\Delta\theta$ ($^\circ$) \\
\midrule
\textbf{1P/Halley}      & 76.713 & 29 & 240\,BCE--1986 & 15  & $\mathbf{+1.4}$ \\
12P/Pons--Brooks         & 70.942 & 9  & 1385--2024     & 16  & $+80.8$ \\
12P/Pons--Brooks         & 70.533 & 3  & 1812--2024     & 16  & $+114.7$ \\
27P/Crommelin            & 27.558 & 3  & 1928--2011     & 42  & $-84.0$ \\
55P/Tempel--Tuttle       & 33.225 & 33 & 901--1998      & 35  & $-128.6$ \\
55P/Tempel--Tuttle       & 33.229 & 19 & 1366--1998     & 35  & $-130.0$ \\
\bottomrule
\end{tabular}

\smallskip
\small\textit{Note.} $N_\mathrm{orb}$: number of orbital periods used to compute $\bar{P}$. $N = \mathrm{nint}(T^{*}/\bar{P})$: nearest integer number of orbits in $T^{*}$. For 12P and 55P, two rows show results for different baselines. Residues of $80^\circ$--$130^\circ$ are comparable to that of Uranus ($-108.3^\circ$), indicating no commensurability.
\end{table}

12P/Pons--Brooks has a nominal period near $T^{*}/16 = 71.94$~yr.
\citet{Meyer2021} identifies six apparitions between 1385 and 2024,
of which four have well-determined dates.
The observed mean period ($70.53$--$70.94$~yr depending on the
baseline) deviates by $\sim 1$~yr from $T^{*}/16$, and individual
periods show a systematic decreasing trend, indicating active
orbital migration rather than stabilization near any commensurable
value.

55P/Tempel--Tuttle has the most extensive computed orbit among the comparison comets, with 33~returns integrated numerically by \citet{Kinoshita2005} from 901 to 1998~CE. Its mean period of $33.225$~yr is far from any commensurable value ($T^{*}/34 = 33.85$~yr or $T^{*}/35 = 32.89$~yr), and the residue of $-128.6^\circ$ is essentially random.

A complementary perspective emerges from the spectral decomposition
of the planetary configuration recurrence spectrum
\citep{BaigetOrts2026c}.
The analysis shows that 16 of the 22 harmonic periods $T^{*}/N$
in the HTC range $[50, 200]$~yr simultaneously coincide with
low-order combinations of Jupiter and Saturn periods
$a \times P_J + b \times P_S$ to within $0.2\%$.
These 16 values constitute dynamically special points where the
$T^{*}$ harmonic structure and the Jupiter--Saturn resonance
network reinforce each other.

Table~\ref{tab:htc_intersections} shows the proximity of each
surveyed HTC to the nearest such intersection.
Halley's mean period lies $0.020$~yr ($0.002\%$ of $T^{*}$) from
the $T^{*}/15$ intersection, which coincides with
$4 \times P_J + P_S = 76.900$~yr to within $0.167$~yr.
12P/Pons--Brooks lies within $1.0$~yr of the $T^{*}/16$
intersection ($6 \times P_J = 71.17$~yr), but at a precision
40--60 times lower than Halley and with a mean period that is
actively migrating away from the intersection value.
27P/Crommelin and 55P/Tempel--Tuttle have no period within
$1.5\%$ of any intersection, placing them outside the network
entirely.

\begin{table}[htbp]
\centering
\caption{Proximity of surveyed HTCs to the nearest
  $T^{*}/N \cap JS$ intersection.}
\label{tab:htc_intersections}
\begin{tabular}{lrrrrrr}
\toprule
Comet & $\bar{P}$ (yr) & Nearest $T^{*}/N$ & $N$ &
  $|\Delta P|$ (yr) & $|\Delta P|/T^{*}$ (\%) \\
\midrule
\textbf{1P/Halley}       & 76.713 & 76.733 & 15 & 0.020 & 0.002 \\
12P/Pons--Brooks (9 app) & 70.942 & 71.938 & 16 & 0.996 & 0.087 \\
12P/Pons--Brooks (3 app) & 70.533 & 71.938 & 16 & 1.405 & 0.122 \\
27P/Crommelin            & 27.558 & 52.318 & 22 & 24.760 & 2.151 \\
55P/Tempel--Tuttle (33)  & 33.225 & 52.318 & 22 & 19.093 & 1.659 \\
55P/Tempel--Tuttle (19)  & 33.229 & 52.318 & 22 & 19.089 & 1.658 \\
\bottomrule
\end{tabular}
\smallskip
\small\textit{Note.} Intersections are periods $T^{*}/N$ that
simultaneously coincide with a combination $a \times P_J + b \times P_S$
to within $0.2\%$. Halley is 40--60 times more precisely positioned
at its intersection than the nearest comparison comet.
\end{table}

\section{Discussion}\label{sec:discussion}

\subsection{Two complementary coupling mechanisms}
\label{subsec:mechanism}

The central result of this work---that Halley's mean orbital period
converges to $T^{*}/15$ despite orbit-to-orbit variations of
$\pm 2$~yr---implies that the perturbations received at successive
perihelion passages are not independent.
Rather, they are modulated by the planetary configuration, which
approximately recurs every $T^{*}$~years.

The dominant perturbers of Halley's orbit are Jupiter and Saturn
\citep{Yeomans1986}.
In $T^{*} = 1{,}151$~yr, Jupiter completes $\sim 97$ orbits and
Saturn $\sim 39$ orbits, both near-integers.
Three independent statistical tests, described in
Sections~\ref{subsec:jupiter} and~\ref{subsec:saturn} below,
show that Jupiter and Saturn act through
\textit{distinct but complementary mechanisms}: Jupiter through
phase-dependent modulation detectable at the perihelion-to-perihelion
level, and Saturn through a distance-amplitude modulation whose
signature is orthogonal to phase.

This two-mechanism picture is physically natural.
Jupiter, at a mean distance of $\sim 5.3$~AU from Halley's perihelion,
is close enough for its angular geometry to impose a directional
perturbation: the sign and magnitude of $\delta P_i$ depend on
whether Jupiter is ahead of or behind Halley in its orbit.

Saturn, at $\sim 9.5$~AU, is sufficiently distant for its angular
position to become secondary; what varies significantly is its
\textit{absolute distance} from Halley, which modulates the amplitude
of the perturbation it delivers regardless of direction.

The dynamical significance of $T^{*}/\bar{P}$ being close to an
integer deserves emphasis.
If $\bar{P} = T^{*}/N$ exactly for integer $N$, then after $N$ orbits
of Halley exactly $T^{*}$~years have elapsed, and the planetary
configuration has returned to a state nearly identical to the initial
one.
The sequence of gravitational impulses that Halley receives at the $N$
perihelion passages of one cycle is therefore approximately repeated
in the next cycle.
If this sequence sums to zero---as the observed cancellation at
$n = 15$ orbits suggests---then the mean period is self-consistently
preserved: the same perturbation pattern recurs, and cancels, in every
subsequent cycle.

This is structurally analogous to the Laplace resonance among Io,
Europa, and Ganymede, where the integer period ratios $1:2:4$ ensure
that the gravitational interactions repeat coherently and their net
effect on the orbital energies cancels over one resonance cycle.
The key property in both cases is not the resonance itself but the
\textit{coherence of phase} it implies: with a non-integer ratio
$T^{*}/\bar{P} = 15.3$, for example, the planetary configuration
would drift between successive Halley cycles, the impulse sequence
would differ each time, and systematic cancellation would have no
reason to occur.
The near-integer value $T^{*}/\bar{P} = 15.004$ is thus not merely
an arithmetic curiosity: it is the condition that makes coherent
cancellation dynamically possible.

A potential concern is the look-elsewhere effect: $T^{*}$ was
identified by optimizing over planetary longitudes, and one may
ask whether the Halley commensurability benefits from the same
implicit degrees of freedom. Three observations mitigate this
concern. First, $T^{*}$ was derived from the planets alone, with
no reference to cometary data; the Halley analysis was performed
independently on a separately defined quantity. Second, the joint
Monte Carlo test ($p = 0.009$) explicitly accounts for the
search over candidate cycles from 100 to 2,000~yr. Third, the
physical mechanism --- Jupiter phase coupling and Saturn distance
modulation --- provides an independent line of evidence that does
not depend on the numerical value of $T^{*}$.

\subsection{Jupiter: phase-dependent modulation}
\label{subsec:jupiter}

I test the prediction that $\delta P_i$ correlates with Jupiter's
orbital configuration at each perihelion passage using two independent
methods.

\textit{Circular-linear phase correlation.}
The circular-linear correlation between $\delta P_i$ and Jupiter's
heliocentric ecliptic longitude $\lambda_J$ at the start of period $i$
yields $R = 0.47$, $p = 0.04$, with a perturbation amplitude of
$\sim 427$~days ($\sim 1.2$~yr).
An independent check using the simplified orbital phases of
\citet{ChirikovVecheslavov1989} gives virtually identical results,
confirming robustness to the choice of planetary position source.

As an independent check, the tidal gravitational impulse
$\Delta E_J^{(i)}$ delivered by Jupiter at each perihelion correlates
with $\delta P_i$ at $r = -0.41$ (permutation $p = 0.027$), with the
physically correct negative sign: a prograde kick shortens the
semi-major axis and the next period.
The amplitude ($\sim 427$~days) is consistent with the phase-correlation
result, confirming that angular position is an adequate proxy for the
perturbation geometry.

\textit{Phase-locked permutation test.}
A third, more stringent test asks whether the specific
\textit{pairing} of observed $\delta P_i$ values with the real
Jupiter--Saturn phase sequence carries information beyond what the
$\delta P$ distribution alone contains.
I fix the 29 real planetary phase values and permute the 29
observed $\delta P_i$ across them, measuring the $R^2$ of the
combined Jupiter--Saturn sinusoidal model for each of $10^6$
permutations.
The observed $R^2 = 0.223$ (Jupiter only) lies at the 96.5th percentile
of the permutation distribution, yielding $p = 0.035$.
This test is immune to the ``cancellation by chance'' critique because
both the observed case and every permutation share identical
cancellation statistics; only the pairing information differs.

Three independent tests thus converge on the same conclusion:
Jupiter's angular configuration at each perihelion predicts the
sign and magnitude of Halley's period deviation, with a
permutation-based $p$-value of $0.027$--$0.035$ across methods.

\subsection{Saturn: distance-amplitude modulation}
\label{subsec:saturn}

Jupiter's amplitude ratio over Saturn in the circular-linear
correlation ($3.5\times$) and in the tidal impulse ($11.2\times$)
both reflect Saturn's greater distance.
Correspondingly, Saturn's \textit{phase} does not predict $\delta P_i$:
the circular-linear correlation with $\lambda_S$ is $R = 0.13$,
$p = 0.77$, and the phase-locked permutation test gives $p = 0.780$.

However, a different and physically motivated metric reveals
Saturn's contribution clearly.
I compute the Halley--Saturn distance $d_S^{(i)}$ at each perihelion
using the discrete Kepler-map approximation of
\citet{ChirikovVecheslavov1989}: the elapsed time between consecutive
perihelion passages determines Saturn's angular position via uniform
circular motion at its mean orbital radius, from which the
instantaneous Halley--Saturn distance follows geometrically.
I then correlate $d_S^{(i)}$ with $|\delta P_i|$ (the magnitude
of the period deviation, sign-agnostic).
The result is $r = -0.496$, confirmed by a $10^6$-permutation test
at $p = 0.0066$: closer Saturn approaches produce systematically
larger period deviations.

Three further tests establish that this is a genuine dynamical
coupling, not a spurious distance trend.

\textit{Sign consistency.}
Among the 10 closest Saturn approaches, positive and negative
$\delta P_i$ values are equally distributed (4 positive, 6 negative;
binomial $p = 0.75$).
Saturn amplifies the perturbation amplitude regardless of direction.
This is the expected signature of a distance-driven mechanism: a
nearby Saturn increases the magnitude of whatever perturbation the
full planetary configuration delivers, without imposing a preferred
direction of its own.

\textit{Random-phase Saturn orbit.}
The key discriminating test is whether the correlation
survives when Saturn's orbit is intact but its phase is shifted
randomly relative to Halley's perihelion dates.
I generate $10^5$ Saturn distance sequences with the same orbital
parameters but a uniformly random phase offset $\phi_0 \in [0, 2\pi)$,
and compute $|r|$ for each.
The resulting $p$-value is $0.133$ --- twenty times larger than
the observed value ($p = 0.007$) --- confirming that the correlation
is specific to the \textit{actual Saturn phase} at Halley's real
perihelion dates, not a generic property of a slowly-varying
outer-planet distance.

The mean $|\delta P_i|$ for the 10 closest Saturn approaches is
$1.34$~yr, compared to $0.76$~yr for the remaining 19 perihelia
(ratio $1.76\times$; Mann--Whitney $p = 0.052$).

Together, these results establish that Saturn modulates the
\textit{amplitude} of Halley's period deviations through its
proximity at each perihelion, while Jupiter modulates their
\textit{direction and magnitude} through its phase.
The two mechanisms are complementary rather than redundant:
Jupiter determines what direction Halley is pushed; Saturn
determines how hard.

\subsection{Cancellation over commensurable cycles}
\label{subsec:cancellation}

A further prediction of the coupling hypothesis is that period
deviations should cancel over multiples of 15~orbits, since the
full planetary configuration --- including both Jupiter's phase
and Saturn's distance --- recurs every $T^{*}$~years.

If the period deviations were a random walk, the expected magnitude
of the cumulative sum after $n$~orbits would be $\sigma\sqrt{n}$,
where $\sigma = 1.25$~yr.
After 15~orbits (one commensurable cycle), the observed cumulative
deviation is $|\sum_{i=1}^{15} \delta P_i| = 0.46$~yr, compared to
the random-walk expectation of $\sigma\sqrt{15} = 4.83$~yr.
The actual sum is only $9.4\%$ of the random-walk expectation.
After 29~orbits (all available data), the cumulative sum is
$< 0.001$~yr, consistent with exact cancellation.

A synthetic-clone test ($10^5$ sequences, both reshuffle and Gaussian
null models) yields a joint $p$-value of $0.012$ for simultaneously
matching the observed cancellation at $n = 15$, the combined $R^2$,
and the predictive advantage of $T^{*}/15$ over the running mean.
This confirms that the pattern of cancellation is not reproducible
by random period sequences with the same marginal statistics.

\begin{figure}[htbp]
  \centering
  \includegraphics[width=0.85\textwidth]{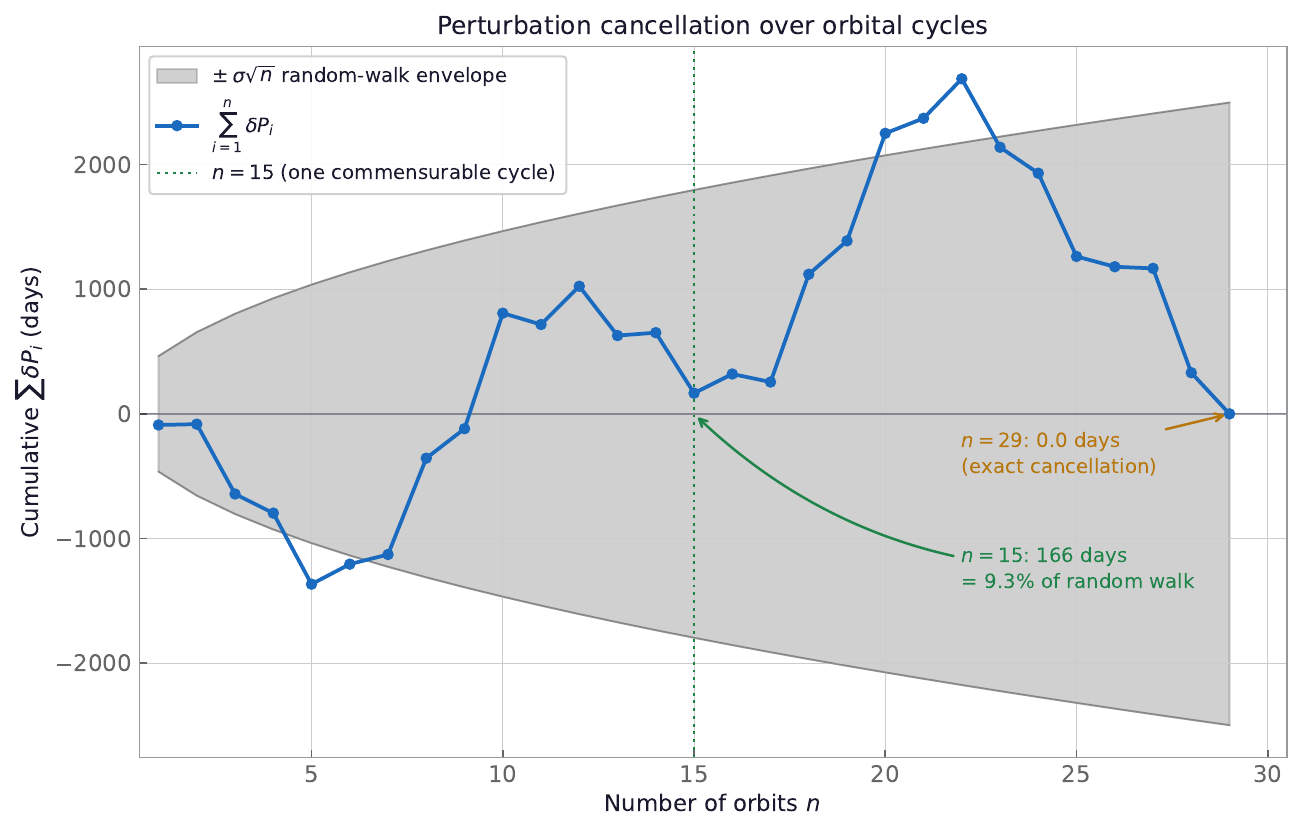}
  \caption{Cumulative sum of period deviations $\sum_{i=1}^{n}\delta P_i$
    (blue line, in days) as a function of the number of orbits $n$.
    The grey shaded band shows the $\pm\sigma\sqrt{n}$ random-walk
    envelope expected if perturbations were independent.
    The green vertical line marks $n = 15$ (one commensurable cycle),
    where the cumulative sum is only $9.4\%$ of the random-walk
    expectation.
    At $n = 29$ (all data), the sum collapses to $< 1$~day,
    consistent with exact cancellation.
    The excursion to $+2{,}500$~days between $n = 15$ and $n = 29$
    reflects the chaotic variability of individual periods;
    the cancellation operates over the full 29-orbit baseline,
    not monotonically.}
  \label{fig:cumulative_perturbation}
\end{figure}

Figure~\ref{fig:cumulative_perturbation} displays the cumulative sum
alongside the $\pm\sigma\sqrt{n}$ random-walk envelope.

\subsection{Chaos and commensurability}\label{subsec:chaos}

The commensurability reported here must be reconciled with the
established chaotic nature of Halley's orbit.
\citet{MunozGutierrez2015} computed the Lyapunov exponent for Halley
directly from N-body integrations, finding a positive value
corresponding to a Lyapunov time of $\tau_L \approx 70$~yr.
This means that orbits initially separated by the observational
uncertainty ($\sim 10^{-6}$~au) diverge by a factor of $e$ over
approximately one orbital period.

At first sight, this appears incompatible with a long-term
commensurability maintained over 29~orbits.
However, short-term chaotic divergence and long-term statistical
regularity are well-known coexisting phenomena in Hamiltonian
dynamics.
The Lyapunov time governs the predictability of the
\textit{instantaneous trajectory}.
The commensurability reported here concerns the \textit{mean period},
a time-averaged quantity governed by approximate integrals of motion
and the large-scale topology of phase space --- a level at which the
Lyapunov exponent is not directly relevant.

An analogy from \citet{MunozGutierrez2015} themselves is
instructive: their Figure~10 shows that 200~particles with initial
conditions indistinguishable from Halley's evolve along widely
different trajectories in $a$, $e$, and $i$, yet the Tisserand
parameter with respect to Jupiter, $T_J$, is approximately conserved
for the ensemble.
The mean orbital period may be constrained by a similar
quasi-invariant associated with the $T^{*}$ quasi-periodicity of
the planetary configuration.

Furthermore, the chaotic dynamics may actively contribute to the
commensurability rather than opposing it.
If the region of phase space near $\bar{P} = T^{*}/15$ constitutes
an attractor in the time-averaged sense---where perturbations from
Jupiter and Saturn cancel over multiples of 15~orbits---then chaotic
diffusion within this region would redistribute the comet's
trajectory without systematically driving the mean period away from
the commensurable value.
The statistical evidence from Sections~\ref{subsec:jupiter},
\ref{subsec:saturn}, and~\ref{subsec:cancellation} supports this picture.

It is worth noting that \citet{ChirikovVecheslavov1989}, in their
pioneering study of Halley's chaotic dynamics, explicitly concluded
that the search for commensurabilities using the ``cyclic method''
is ``totally inapplicable here'' precisely because the motion is
chaotic.
The present result demonstrates that this conclusion, while correct
regarding the instantaneous trajectory, does not extend to the
long-term mean period --- a quantity that their framework was not
designed to detect.

\subsection{N-body stability maps}\label{subsec:nbody}

To test whether $T^{*}/15$ corresponds to a dynamically preferred
region in phase space, I performed three N-body experiments using
REBOUND \citep{Rein2012} with the IAS15 integrator
\citep{ReinSpiegel2015}, initialized from JPL Horizons elements at
J2000 for all eight planets, with test particles treated as massless.

\textit{Experiment 1: period stability map.}
500~massless test particles with Halley's orbital elements
($e, i, \Omega, \omega$ fixed) but initial periods uniformly
distributed in $[70, 84]$~yr were integrated for $10{,}000$~yr.
The stability score (period drift $+$ diffusion) shows no minimum
near $T^{*}/15 = 76.73$~yr.
The minimum appears near $T^{*}/16 = 71.94$~yr
($P_\mathrm{min} = 71.26$~yr), and the Pearson correlation between
score and $|P_\mathrm{init} - T^{*}/15|$ is $r = -0.12$, indicating
no significant preference for the $T^{*}/15$ region.

\textit{Experiment 2: Saturn phase control.}
The experiment was repeated with Saturn's initial mean anomaly shifted
by $\pi$.
The result is structurally identical: minimum score near $T^{*}/16$
($P_\mathrm{min} = 71.46$~yr), $r = -0.12$.
The structure is insensitive to Saturn's phase, indicating that it
reflects the arithmetic density of low-order commensurabilities
in the $[70, 84]$~yr range rather than a specific dynamical effect.

\textit{Experiment 3: phase-space map over $(P, M_0)$.}
A 2D grid of $30 \times 10 = 300$~particles varying both initial
period and initial mean anomaly was integrated for $20{,}000$~yr.
The minimum score averaged over all initial phases occurs again near
$T^{*}/16$ ($P = 71.45$~yr), and no localized minimum appears at
$(P \approx T^{*}/15,\, M_0 \approx 38^\circ)$, the actual
initial conditions of Halley.

Taken together, these experiments indicate that $T^{*}/15$ is not a
generic stability island for Halley-type orbits: the commensurability
does not arise because the region is dynamically protected for
arbitrary particles with Halley's shape elements.
This is fully consistent with the uniqueness result of
Section~\ref{subsec:uniqueness}: the commensurability requires
Halley's specific combination of orbital parameters and dynamical
history, not merely the value of its period.
The question of what dynamical mechanism selects $T^{*}/15$ over
neighbouring commensurabilities for this particular orbit remains
open for future N-body investigation with longer integration times
and varying $(e, i)$.

\subsection{Why Halley and not others?}\label{subsec:uniqueness}

\begin{figure}[htbp]
  \centering
  \includegraphics[width=\textwidth]{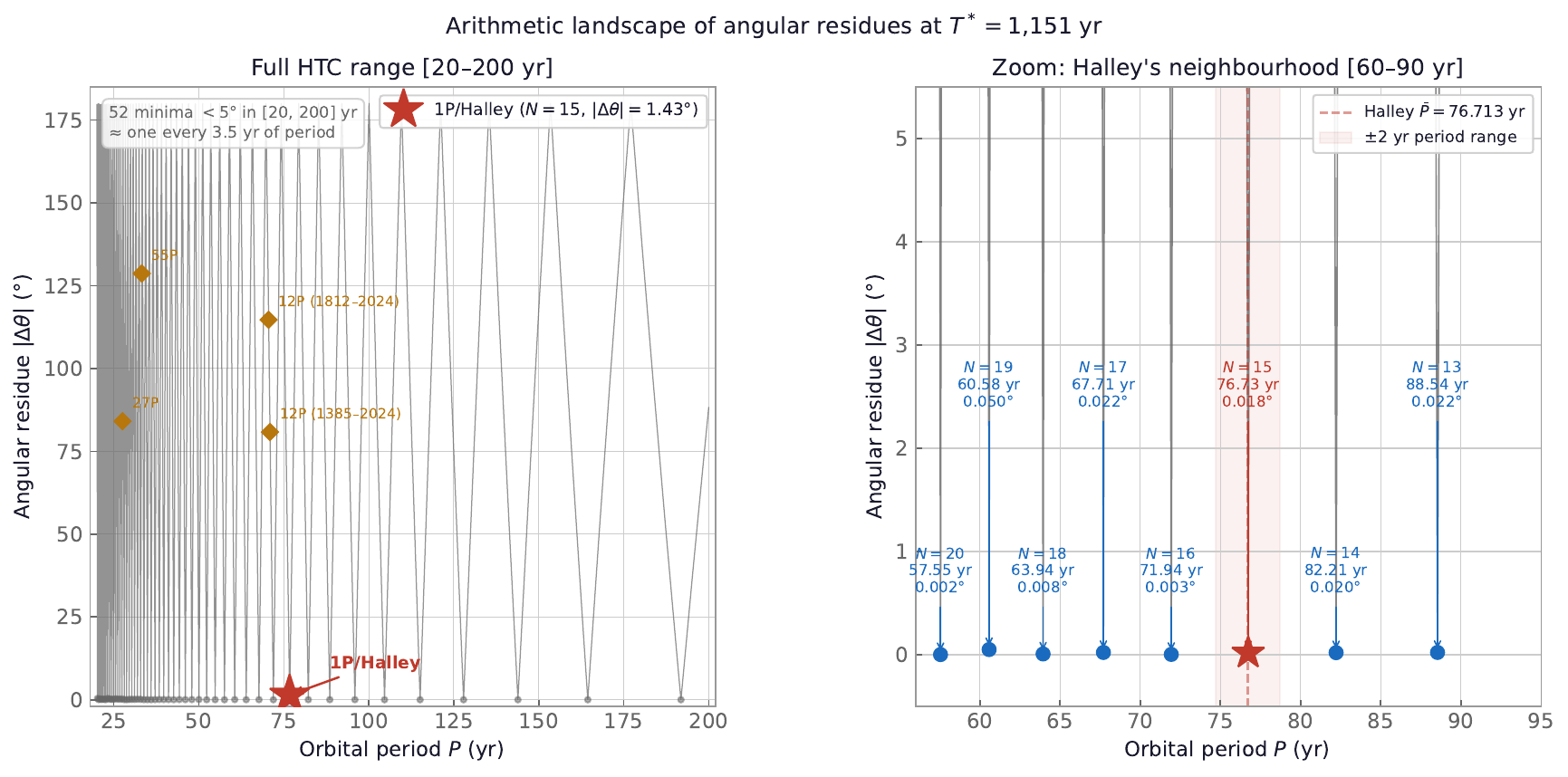}
  \caption{Arithmetic landscape of angular residues $|\Delta\theta|$
    at $T^{*} = 1{,}151$~yr as a function of orbital period.
    \emph{Left panel}: full Halley-type comet range $[20, 200]$~yr.
    Grey dots mark local minima below $5^\circ$ (52~total);
    the red star marks 1P/Halley; orange diamonds mark other HTCs.
    \emph{Right panel}: zoom on Halley's neighbourhood $[57, 94]$~yr.
    The landscape is densely populated: $T^{*}/15$ is not an
    arithmetically isolated or uniquely deep minimum, demonstrating
    that Halley's convergence to this value is a dynamical result
    rather than an arithmetic accident.}
  \label{fig:arithmetic_landscape}
\end{figure}

The absence of commensurability in other HTCs
(Section~\ref{subsec:other_htcs}) demonstrates that the coupling
is not a generic property of Jupiter--Saturn perturbations on
cometary orbits.
Rather, it requires a specific combination of orbital parameters.
Several factors may distinguish Halley:

\begin{enumerate}

\item \textit{Low-order commensurability.}
Halley's $N = 15$ is the lowest integer multiplier among the comets
examined: 12P/Pons--Brooks requires $N = 16$, 55P/Tempel--Tuttle
$N = 34$--35, and 27P/Crommelin $N = 41$--42.
Lower-order resonances are generally stronger and more robust to
perturbations \citep{MurrayDermott1999}.
However, the small difference between $N = 15$ and $N = 16$
(12P/Pons--Brooks) is insufficient to explain alone the
contrast in residues ($1.4^\circ$ versus $80^\circ$--$115^\circ$).
The decisive factor is discussed in point~5 below: $T^{*}/15$
coincides with a Jupiter--Saturn combination to $0.015\%$,
while $T^{*}/16$ does so only to $1.1\%$, forty times less precisely.

\item \textit{Retrograde orbit.}
Halley's retrograde motion ($i = 162.3^\circ$) produces shorter
synodic periods with the giant planets than a prograde orbit of
the same period would.
For a retrograde orbit, the synodic period with planet $p$ is
$P_\mathrm{syn} = (1/P_p + 1/P_H)^{-1}$ rather than the prograde
$(1/P_p - 1/P_H)^{-1}$.
This increases the frequency of perturbative encounters, potentially
strengthening the phase-averaging effect over $T^{*}$.
Among the comparison comets, only Tempel--Tuttle shares a similar
inclination ($i = 162.5^\circ$), but at a much shorter period
($N = 34$--35).

\item \textit{Orbital longevity.}
Halley has a median survival time of $\sim 3.2 \times 10^{5}$~yr
\citep{MunozGutierrez2015}.
If Halley has been in its current orbit for several $T^{*}$ cycles
($> 10{,}000$~yr), there has been ample time for dynamical relaxation
toward commensurability.
Comets with shorter dynamical lifetimes or more recent captures may
not have had time to stabilize.

\item \textit{Dense arithmetic landscape.}
A sweep of the angular residue $|\Delta\theta(T^{*}, P)|$ over all
periods in $[20, 200]$~yr reveals 52~local minima below $5^\circ$.
In Halley's immediate neighbourhood, minima with comparable residues
occur at $P \approx 71.9$~yr ($N=16$) and $P \approx 82.2$~yr
($N=14$), separated from $T^{*}/15$ by only $4.8$ and $5.5$~yr
respectively.
Halley's period is not arithmetically special: the landscape near
$76.7$~yr is densely populated with minima of equal depth.
What requires explanation is not \textit{that} $T^{*}/15$ is a
deep minimum, but \textit{why} Halley converges to this particular
minimum while no other HTC does so for any of the equally available
neighbouring ones.
The commensurability is a dynamical result, not an arithmetic accident
(Figure~\ref{fig:arithmetic_landscape}).

\item \textit{Position within the $T^{*}$--JS intersection network.}
A fifth distinguishing property follows from the spectral analysis
of \citet{BaigetOrts2026c}.
Among the 16 dynamically special periods $T^{*}/N \cap JS$ in the
HTC range, Halley occupies its intersection with a precision of
$0.002\%$ of $T^{*}$ --- 40--60 times more precisely than
12P/Pons--Brooks, the only other HTC with a period near any
intersection.
The uniqueness of Halley is therefore not merely that its period
happens to be close to $T^{*}/15$, but that it occupies with
extraordinary precision the exact point where the $T^{*}$ harmonic
structure and the Jupiter--Saturn resonance network coincide.\label{item:network}

\end{enumerate}

\subsection{Relationship to the Uranus anomaly}\label{subsec:uranus}

The quasi-commensurability reported in \citet{BaigetOrts2026a}
encompasses seven of the eight planets in the Solar System.
The sole exception is Uranus (residue $-108.3^\circ$,
$N \approx 13.7$ orbits), which is also the only planet known to
have suffered a catastrophic giant impact during the early Solar
System \citep{Kegerreis2018}.
Neptune---a near-twin in mass and radius that did not experience
a comparable event---participates with high precision
(residue $-5.2^\circ$, $N = 7$ orbits).
The present result adds a new dimension: a cometary body participates
with even higher precision than any planet
(Halley residue $+1.4^\circ$, Table~\ref{tab:residues}).

This strengthens the interpretation that the 1{,}151-year quasi-period
is not merely an arithmetic coincidence among planetary periods,
but a dynamical property of the Solar System that actively organizes
the orbits of bodies subject to Jupiter--Saturn perturbations.
Uranus's exclusion and Halley's inclusion are both consequences of
the same underlying dynamics, viewed from opposite ends: Uranus was
perturbed \textit{out of} commensurability by the giant impact, while
Halley was drawn \textit{into} it by the very forces that define the
cycle.

\subsection{Rolling prediction test}\label{subsec:rolling}

A defining difference between a genuine dynamical pattern and
numerical coincidence is \textit{predictive power}: a coincidence
can fit past data but cannot predict future observations better
than the data's own statistics.

Starting from the first $n$ perihelion dates ($n \geq 3$), I predict
the $(n+1)$-th perihelion date using three methods: (1)~the running
mean period from the $n-1$ observed orbital periods; (2)~the fixed
commensurable period $T^{*}/15 = 76.733$~yr, derived entirely from
the planetary quasi-period with no Halley data; and (3)~the last
observed orbital period (a naïve predictor).

The results (Table~\ref{tab:rolling}) show that $T^{*}/15$
outperforms the running mean.
The predictive advantage of $T^{*}/15$ must be interpreted with
care: a synthetic-clone test with $10^5$ random period sequences
shows that $\sim 80\%$ of random sequences also ``prefer''
$T^{*}/15$ over their own running mean as a predictor, due to the
small numerical proximity between $\bar{P}$ and $T^{*}/15$
(7.4~days).
The rolling test is therefore \textit{consistent with} the
commensurability but is not by itself sufficient evidence for it;
the phase-coupling and distance-modulation results of
Sections~\ref{subsec:jupiter} and~\ref{subsec:saturn} provide the
primary statistical evidence.
The results are summarized in Table~\ref{tab:rolling}: RMS error
of 472~days for $T^{*}/15$ versus 494~days for the running mean
and 594~days for the last-period predictor.
\begin{table}[htbp]
\centering
\caption{One-step-ahead prediction test: summary statistics over
  27~forecasts.}
\label{tab:rolling}
\begin{tabular}{lrrr}
\toprule
Metric & Running mean & $T^{*}/15$ fixed & Last period \\
\midrule
Mean signed error (days)  & $-1.9$  & $+4.4$  & $+12.5$ \\
Mean $|$error$|$ (days)   & $395.7$ & $374.2$ & $545.7$ \\
Median $|$error$|$ (days) & $301.1$ & $299.4$ & $566.5$ \\
RMS error (days)          & $493.7$ & $471.8$ & $594.5$ \\
RMS error (years)         & $1.35$  & $1.29$  & $1.63$  \\
\bottomrule
\multicolumn{4}{p{0.85\textwidth}}{\small\textit{Note.} The $T^{*}/15$
predictor uses a fixed period derived from the planetary quasi-period,
with no Halley data. Source code for this test is publicly available.} \\
\end{tabular}
\end{table}

\begin{figure}[htbp]
  \centering
  \includegraphics[width=\textwidth]{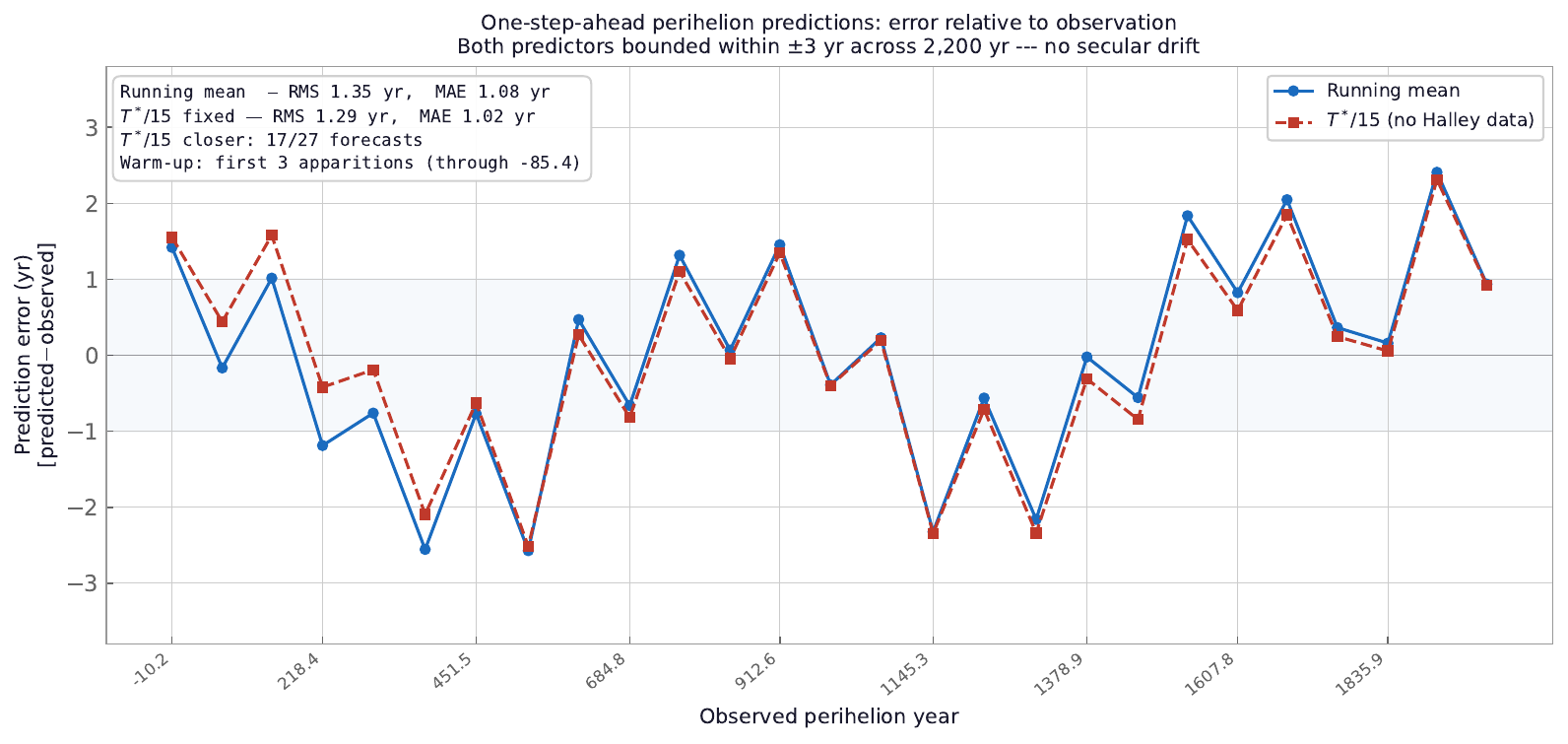}
  \caption{One-step-ahead perihelion prediction errors across
    2{,}225~yr of Halley observations (27~forecasts, warm-up:
    first 3~apparitions through $-86.4$).
    Blue circles: error of the running mean of all previously
    observed periods.
    Red squares: error of the fixed period $T^{*}/15 = 76.733$~yr,
    derived solely from the planetary quasi-period with no Halley
    data.
    The grey band marks $\pm 1$~yr for reference.
    Both predictors remain bounded within $\pm 3$~yr throughout
    the 2{,}200-yr baseline with no secular drift, despite
    orbit-to-orbit period variability of $\pm 2$~yr.
    RMS errors: running mean $1.35$~yr; $T^{*}/15$ fixed
    $1.29$~yr (17 of 27~forecasts, 63\%).}
  \label{fig:rolling_prediction}
\end{figure}

\subsection{Chaos at the orbit scale, stability at the millennium
scale}\label{subsec:chaos_stability}

A natural question raised by any long-term statistical analysis of
Halley's apparitions is whether the apparent regularity of its
mean return interval over two millennia is physically consistent
with a formally chaotic orbit.
The orbit of comet 1P/Halley is chaotic: \citet{ChirikovVecheslavov1989}
demonstrated this through the Kepler-map formalism, and
\citet{MunozGutierrez2015} quantified a Lyapunov time of $\sim 70$~yr
--- shorter than one orbital period --- implying that individual
trajectory predictions lose accuracy on a timescale comparable to
a single revolution.
Yet the same historical record that establishes the chaotic nature
of the orbit also documents 29 perihelion passages spanning
2{,}225~years with a mean period stable to 7.4~days.
The apparent contradiction dissolves once the relevant timescales
are carefully distinguished.

The Lyapunov time governs the exponential divergence of
\emph{individual trajectories}: a small uncertainty in the initial
conditions of any single orbit grows by a factor $e$ every
$\sim 70$~yr.
It does not, however, constrain the long-term \emph{mean period},
which is an ensemble average over many successive perturbation
events rather than a property of any single orbit.
That individual chaos need not preclude long-term mean regularity
is a well-established feature of nonlinear dynamics
\citep{MurrayDermott1999}, and recent work explicitly identifies
the orbital evolution of comet Halley as a case of
\emph{confined} chaos---strong short-term divergence coexisting
with bounded long-term excursions of the orbital elements
\citep{MunozGutierrez2015}.

The commensurability result of Section~\ref{subsec:halley_result}
provides a direct, quantitative expression of this confinement.
The perturbations delivered by Jupiter and Saturn at each
perihelion passage are not independent random kicks; they are
correlated with the planetary configuration at the moment of
closest approach, and that configuration recurs every
$T^{*} = 1{,}151$~yr \citep{BaigetOrts2026a}.
The perturbation-cancellation test of
Section~\ref{subsec:cancellation} quantifies the consequence:
after 15~orbits (one complete $T^{*}$ cycle), the cumulative
period deviation is only $9.4\%$ of the random-walk expectation.
A purely random perturbation sequence---one that would be
consistent with naive chaos---would produce $\sim 100\%$;
the observed value indicates that the perturbations are
systematically correlated over the $T^{*}$ baseline, not
randomly distributed.

Figure~\ref{fig:rolling_prediction} illustrates the same phenomenon
from an observational perspective.
Starting from only the first three perihelion passages,
one-step-ahead forecasts of each subsequent apparition are computed using two
predictors: the running mean of all previously observed periods
(a purely data-driven approach), and the fixed period
$T^{*}/15 = 76.733$~yr derived solely from the planetary
quasi-period with no Halley data whatsoever.
Both predictors track the observed dates across the full
2{,}200-year baseline with errors bounded within $\pm 3$~yr and
no secular drift, despite the orbit-to-orbit
variability of $\pm 2$~yr that expresses the chaotic dynamics.
The fixed planetary predictor achieves RMS~$= 1.29$~yr versus
RMS~$= 1.35$~yr for the running mean---a body whose chaotic
dynamics have been thoroughly documented is predicted more
accurately by a number derived from the planets than by its own
observational history.

This behaviour is physically expected under the commensurability
hypothesis.
Orbit-to-orbit chaos scrambles the short-term period sequence,
but the long-term mean is anchored to $T^{*}/15$ by the
Jupiter--Saturn coupling demonstrated in
Sections~\ref{subsec:jupiter} and~\ref{subsec:saturn}.
The perturbations are correlated noise, not white noise, and their
correlation timescale is set by the planetary quasi-period.
Long-term mean stability is therefore not despite the chaotic
dynamics but a \emph{consequence} of the same planetary forces
that cause the individual-orbit variability: Jupiter and Saturn
perturb each orbit, but the net effect over one $T^{*}$ cycle
nearly cancels by construction.

In summary, the historical record of comet Halley over two
millennia is fully consistent with its known chaotic dynamics.
The stability is not of individual orbits---those remain
unpredictable beyond one Lyapunov time---but of the long-term mean
period, which is constrained by the 1{,}151-year quasi-period of
the Solar System.
Regularity at the century-to-millennium scale and chaos at the
orbital scale are not contradictory but complementary expressions
of the same underlying Jupiter--Saturn dynamics.


\section{Conclusions}\label{sec:conclusions}
 
\begin{enumerate}
 
\item \textbf{Commensurability.}
Comet 1P/Halley completes $15.004$ orbits in one $T^{*} = 1{,}151$-year
planetary quasi-period, corresponding to an angular residue of
$+1.43^\circ$ --- the smallest of any Solar System body examined with
this method, smaller than those of all seven participating planets.
The long-term mean period ($\bar{P} = 76.713$~yr, from 29~observed
orbital periods spanning 2{,}225~years) differs from the exactly
commensurable value $T^{*}/15 = 76.733$~yr by only 7.4~days
($0.026\%$).
A Monte Carlo joint test gives $p = 0.009$.
 
\item \textbf{Uniqueness among HTCs.}
No other Halley-type comet with multiple observed apparitions
participates in the commensurability.
12P/Pons--Brooks, 55P/Tempel--Tuttle, and 27P/Crommelin all exhibit
angular residues of $80^\circ$--$130^\circ$, comparable to that of
Uranus ($-108.3^\circ$), the sole planetary non-participant.
Halley is dynamically unique among known periodic comets.
 
\item \textbf{Jupiter: phase-dependent modulation.}
Three independent tests establish that Jupiter's angular position at
each perihelion passage predicts the sign and magnitude of the period
deviation $\delta P_i$:
circular-linear correlation $R = 0.47$ ($p = 0.04$);
direct gravitational impulse correlation $r = -0.41$
(permutation $p = 0.027$, physically correct sign);
phase-locked permutation test $p = 0.035$.
The perturbation amplitude is $\sim 427$~days ($\sim 1.2$~yr).
 
\item \textbf{Saturn: distance-amplitude modulation.}
Saturn's distance from Halley at perihelion predicts the amplitude
of the period deviation regardless of sign:
$r(|\delta P_i|,\, d_S^{(i)}) = -0.496$
(permutation $p = 0.007$; sign test $p = 0.75$, confirming no
directional preference).
A random-phase Saturn orbit fails to reproduce the correlation
($p = 0.133$, versus $p = 0.007$ for the real orbit), establishing
that the effect is specific to Saturn's actual phase at Halley's
perihelion dates and not a spurious consequence of its slowly-varying
distance.
The mean $|\delta P_i|$ for the 10 closest Saturn approaches is
$1.76\times$ larger than for the remaining 19 perihelia
(Mann--Whitney $p = 0.052$).
 
\item \textbf{Complementary mechanisms.}
Jupiter and Saturn act through distinct and complementary mechanisms.
Jupiter, at $\sim 5.3$~AU, is close enough to impose a directional
perturbation via its angular geometry.
Saturn, at $\sim 9.5$~AU, is sufficiently distant that its angular
position is secondary; its absolute distance modulates perturbation
amplitude without a preferred direction.
The Jupiter-to-Saturn perturbation amplitude ratio is $3.5\times$
(phase method) to $11.2\times$ (impulse method), consistent with
Jupiter's dominant role established by \citet{ChirikovVecheslavov1989}.
 
\item \textbf{Perturbation cancellation.}
After 15~orbits (one commensurable cycle), the cumulative period
deviation $|\sum_{i=1}^{15} \delta P_i| = 0.46$~yr is only $9.4\%$
of the random-walk expectation $\sigma\sqrt{15} = 4.83$~yr.
After 29~orbits (all available data), the cumulative sum is
$< 0.001$~yr, consistent with exact cancellation.
A synthetic-clone test ($10^5$ sequences) yields a joint $p = 0.012$
for simultaneously matching the observed cancellation, $R^2$, and
overall agreement with the commensurable period.
 
\item \textbf{Chaos and commensurability coexist.}
Halley's orbit has a Lyapunov time of $\sim 70$~yr
\citep{MunozGutierrez2015}, shorter than one orbital period.
The commensurability concerns the \textit{mean period} --- a
time-averaged quantity insensitive to chaotic divergence of individual
trajectories.
The conclusion of \citet{ChirikovVecheslavov1989} that commensurability
searches are inapplicable to Halley's chaotic orbit is correct for the
instantaneous trajectory but does not extend to the long-term mean
period, a level their framework was not designed to probe.
 
\item \textbf{Relationship to the Uranus anomaly.}
Halley's angular residue ($+1.4^\circ$) is smaller than those of all
seven planets participating in $T^{*}$, while Uranus ($-108.3^\circ$)
--- the only planet subjected to a catastrophic giant impact ---
is the sole non-participant among both planets and the HTCs examined.
Uranus's exclusion and Halley's inclusion are consequences of the same
dynamics viewed from opposite ends: Uranus was perturbed
\textit{out of} commensurability; Halley was drawn \textit{into} it
by the forces that define the cycle.
 
\item \textbf{The 1,151-year quasi-period as a dynamical organizer.}
The commensurability of Halley's mean period with $T^{*}$, coupled
with the statistical evidence for Jupiter--Saturn coupling, strengthens
the interpretation that the 1{,}151-year quasi-period is not merely an
arithmetic coincidence among planetary orbital periods, but a dynamical
property of the Solar System that actively organizes the orbits of
bodies subject to Jupiter--Saturn perturbations.
Whether this organizing role extends to other short-period comets,
asteroids in mean-motion resonance with Jupiter, or other minor-body
populations is an open question for future investigation.

\item \textbf{Position within the $T^{*}$--JS intersection network.}
The harmonic periods $T^{*}/N$ that simultaneously coincide with
Jupiter--Saturn combinations $a \times P_J + b \times P_S$ define
16~dynamically special points in the HTC period range.
Halley's mean period lies $0.020$~yr from the $T^{*}/15$ intersection
($0.002\%$ of $T^{*}$), 40--60 times more precisely than
12P/Pons--Brooks near $T^{*}/16$.
Crommelin and Tempel--Tuttle have no period within $1.5\%$ of any
intersection.
This suggests that participation in the $T^{*}$ quasi-commensurability
requires not merely proximity to a $T^{*}/N$ value, but occupation
of a point where $T^{*}$ harmonics and Jupiter--Saturn resonances
mutually reinforce.

\end{enumerate}

\section*{Data Availability}

The perihelion dates used in this work are compiled from \citet{Yeomans1986} and publicly available catalogs. Complete source code for the statistical analysis is available at \url{https://github.com/carbaior/halley_1151} \citep{code2026}.

\section*{Acknowledgments}
The author acknowledges the use of historical perihelion records
compiled by D.~K.~Yeomans and collaborators at the Jet Propulsion
Laboratory. Orbital elements and computed return dates for
Halley-type comets were obtained from the JPL Small-Body Database
and the Minor Planet Center. The author thanks the developers of
the NumPy \citep{vanderWalt2011}, Matplotlib \citep{Hunter2007},
and REBOUND \citep{Rein2012} libraries, and the IAS15 integrator
\citep{ReinSpiegel2015}.


\end{document}